\documentclass[page-classic]{epl2} 

\title{Creating artificial magnetic fields for cold atoms by photon-assisted tunneling}
\shorttitle{Creating artificial magnetic fields for cold atoms}

\author{Andrey R. Kolovsky\inst{1,2}}
\shortauthor{A.R.Kolovsky}

\institute{
  \inst{1} Kirensky Institute of Physics, 660036 Krasnoyarsk, Russia \\
  \inst{2} Siberian Federal University, 660041 Krasnoyarsk, Russia
}
\pacs{03.75.Lm}{BEC in periodic potentials, solitons}
\pacs{73.43-f}{Quantum Hall effects}
\pacs{05.45.-a}{Nonlinear dynamics and chaos}

\abstract{This paper proposes a simple setup for introducing an artificial magnetic field for neutral atoms in 2D optical lattices. This setup is based on the phenomenon of photon-assisted tunneling and involves a low-frequency periodic driving of the optical lattice. This low-frequency driving does not affect the electronic structure of the atom and can be easily realized by the same means which employed to create the lattice. We also address the problem of detecting this effective magnetic field. In particular, we study the center of mass wave-packet dynamics, which is shown to exhibit certain features of cyclotron dynamics of a classical charged particle.}

\begin{document}

\maketitle

\section{Introduction}

A major motivation of the current research with cold atoms in optical lattices is the prospect of simulating solid state physics. However, to have a full access to solid state-like phenomena in these systems (which can be considered as artificial crystals) artificial electric and magnetic fields must be introduced. In present days experiments an external electric field is routinely mimicked by accelerating the optical lattice \cite{Daha96,Mors01,Sias08}, by using the gravitational force \cite{Ande98,Gust08,Albe09} and by a combination of gravitational and levitational forces \cite{Hall10}. The case of artificial magnetic field is more difficult to achieve in laboratory realizations because it requires a setup where the atomic wave function acquires a finite phase when the atom tunnels along a closed path on the lattice. 

Such a setup was suggested in the seminal paper \cite{Jaks03}, where the authors used two independent 2D optical lattices for two different internal atomic states, which are coupled by additional Raman lasers. The Rabi transition between internal states induces hopping of the atom between nearest lattice sites, where the required phase accumulation is achieved by using a special geometry for the Raman beams. This idea was further developed in Ref.~\cite{Muel04,Oster05}. It was shown that one can also introduce non-Abelian gauge potentials by using the Raman-laser technique. For atoms in a harmonic trap (no lattice) the Raman scheme with 3 spin states of the $F=1$ electronic ground state of $^{87}$Rb atom has been recently realized \cite{Lin09}. In this setup the magnitude of the effective magnetic field is defined by the gradient of the real magnetic field which splits the $F=1$ level into Zeeman sublevels. We note that the cited experiment also reported a drawback of the Raman-laser based techniques. Namely, the spontaneous emission from Raman beams kicks atoms out of the trap, thus causing the population decay.

A different method of introducing artificial magnetic field was proposed in Ref.~\cite{Sore05}. Unlike the setups of Ref.~\cite{Jaks03,Muel04,Oster05,Lin09} it does not rely on the internal atomic structure -- the magnetic field being mimicked by an oscillating quadrupole potential at frequency $\omega$,  together with a periodic modulation of the hopping matrix elements. Here, the magnitude of the effective field is defined by the strength of the quadrupole potential, measured in units of $\hbar\omega$. Unfortunately,  because of a number of approximations involved, this setup does not ensure exact equivalence with a real magnetic field.

In the present work we suggest another, alternative method of creating an artificial magnetic field for neutral atoms in a 2D optical lattice. The method is based on the phenomenon of photon-assisted tunneling, which is well studied in the case of (quasi) 1D lattices \cite{Sias08,Albe09,Hall10}. It has the great advantage over the above discussed setups that it does not involve Raman lasers and, yet, ensures complete equivalence with a charged particle in the real magnetic field.

The problem of creating artificial magnetic fields for neutral atoms is closely related to the detection problem. Here one addresses magnetic field effects, which can be observed by using detection techniques of cold atoms physics. In particular, it was argued in the already cited paper \cite{Jaks03} that the effective magnetic field produces a specific  interference pattern for the atomic density (see Fig.~\ref{f3} below), which carries information about the field magnitude. We revisit here the problem of detecting an artificial magnetic field. In particular, we discuss a variety of interference patterns developed by the atomic wave function in the course of time and study the center of mass wave-packet dynamics, which exhibits certain features of the cyclotron dynamics of a classical particle. 

\section{Photon-assisted tunneling and effective magnetic field} 

The phenomenon of photon assisted tunneling refers to a quantum particle (an atom) in a 1D lattice, which is subject to DC and AC fields.\footnote{See Ref.~\cite{Sias08,Albe09,Hall10} for experimental studies and Ref.~\cite{83,82} for relevant theoretical analysis, which also includes the case of interacting atoms.} 
Using the tight-binding approximation (which will be our theoretical framework from now on) the system Hamiltonian reads 
\begin{equation}
\label{1}
H= -\frac{J}{2} \sum_m (|m+1\rangle \langle m | + h.c.) 
 +a[F+F_\omega\cos(\omega t+\phi)] \sum_m |m \rangle m \langle m | \;,
\end{equation}
where $|m\rangle$ are the Wannier states, $a$ the lattice period and $J$ the hopping matrix element. For vanishing DC and AC fields the egenfunctions of (\ref{1}) are extended Bloch waves with dispersion relation $E(\kappa)=-J\cos(a\kappa)$. If $F\ne0$ the eigenfunctions become localized Wannier-Stark states with an equidistant spectrum with level spacing $\Delta E=dF\equiv\hbar\omega_B$. A periodic driving of the system at the frequency $\omega$ which matches the Bloch frequency $\omega_B$ couples these Wannier-Stark states into extended quasienergy states with dispersion relation $E(\kappa)=-\widetilde{J}\cos(a\kappa)$, where $\widetilde{J}=J{\cal J}_1(F_\omega/F)$ and ${\cal J}_1(x)$ is the Bessel function of the first kind.\footnote{In general case  the photon-assisted tunneling takes place at integer ration $\omega_B/\omega=n$, where $\widetilde{J}=J{\cal J}_n(aF_\omega/\hbar\omega)$.  It may also happen at rational $\omega_B/\omega$, if the Hamiltonian (\ref{1}) includes hopping to the next to nearest-neighbor sites.} Moreover, if 
\begin{equation}
\label{1c}
aF\gg J
\end{equation}
the Bloch period is small in comparison with the characteristic tunneling time and the dynamics of the system (\ref{1}) can be described in terms of the effective Hamiltonian
\begin{equation}
\label{2}
\widetilde{H}= -\frac{\widetilde{J}}{2} \sum_m \left(|m+1\rangle \langle m | e^{i\phi} + h.c.\right) 
\;, \quad \widetilde{J}=J{\cal J}_1(F_\omega/F) \;,
\end{equation}
as already shown in detail in Ref.~\cite{82}. The additional parameter $\phi$ in (\ref{2}), which was actually overlooked in early studies of photon-assisted tunneling, is the phase difference between the Bloch and field oscillations. We mention that  a possibility of  experimental control over the parameter $\phi$ has been recently demonstrated in Ref.~\cite{Hall10}.
\begin{figure}
\center
\includegraphics[width=10cm, clip]{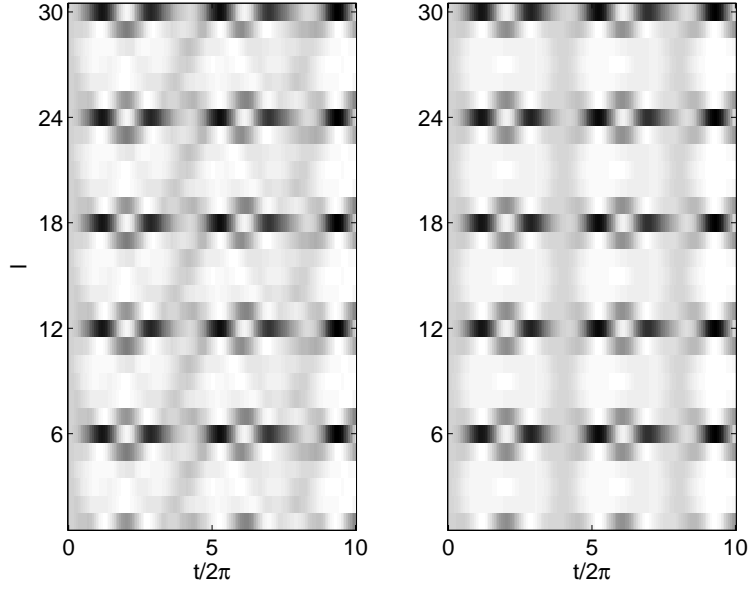}
\caption{Dynamic of the driven system (\ref{3}), left, as compared to dynamic of the effective system (\ref{4}), right. Parameters are $J_x=0.4$, $J_y=2$, $F=1$, $F_\omega=0.4085 F$ (hence $\widetilde{J}_y=0.4$), and $\alpha=1/6$.}
\label{f3}
\end{figure}

Now we consider the quantum particle in a 2D lattice. The lattice is tilted and driven in the $y$ direction (index $m$) and we assume  the phase of the AC field to vary linearly along the $x$ direction (index $l$),
\begin{eqnarray}
\nonumber
H= -\frac{J_x}{2} \sum_{l,m} \left(|l+1,m\rangle \langle l,m | + h.c.\right)
 -\frac{J_y}{2} \sum_{l,m} \left(|l,m+1\rangle \langle m |  + h.c.\right)            \\
\label{3}
+a\sum_{l,m} [F+F_\omega \cos(\omega t-2\pi\alpha l)] |l,m \rangle m \langle l,m | \;.
\end{eqnarray}
Repeating the above arguments for the 1D lattice it is easy to show that this setup realizes the effective Hamiltonian 
\begin{equation}
\label{4}
\widetilde{H}= -\frac{J_x}{2} \sum_l \left(|l+1,m\rangle \langle l,m | + h.c.\right)
-\frac{\widetilde{J}_y}{2} \sum_m \left(|l,m+1\rangle \langle m | e^{i2\pi\alpha l} + h.c.\right) \;,
\end{equation}
which coincides with the tight-binding Hamiltonian of a charged particle subject to a magnetic field.\footnote{For a charged partical the parameter $\alpha=eBa^2/hc$, where $e$ is the charge and $B$ the magnetic field magnitude.}

To check validity of the approximation  involved in the transition from (\ref{3}) to (\ref{4}) we simulate the dynamics of both systems for a finite lattice size, $1\le l,m \le L$, periodic boundary conditions, and uniform initial wave function, $|\Psi(t=0)\rangle=\sum_{l,m} \psi_{l,m}| l,m \rangle$, $\psi_{l,m}=1/L$. It is easy to show that in this case the wave function remains uniform along the $y$ direction. Thus, following Ref.~\cite{Jaks03}, we consider the quantity $n_l(t)=\sum_m |\psi_{l,m}(t)|^2$.  Time evolution of the density $n_l(t)$ calculated on the basis of Eq.~(\ref{4}) for $J_x=\widetilde{J}_y=0.4$ and $\alpha=1/6$,  is shown in the right panel in Fig.~\ref{f3}. It is seen that the density develops a periodic pattern with the spatial period given by $1/\alpha$ and characteristic time period defined by the cyclotron frequency $\omega_c=2\pi\alpha J/\hbar$. The left panel in Fig.~\ref{f3} shows the dynamics of the system (\ref{3}) for the same $\alpha=1/6$ and $J_x=0.4$, and the other parameters adjusted to have $\widetilde{J}_y=J_x$, namely, $F=1$, $F_\omega=0.4085F$, and $J_y=2$. A comparison of these results indicates that the suggested setup indeed introduces an effective magnetic field with the magnitude defined by the phase gradient of the driving force. We note that to observe a difference between the figures we intentionally choose a relatively small $F$, -- for a larger $F$ the figures would be indistinguishable.
\begin{figure}
\center
\includegraphics[width=10cm, clip]{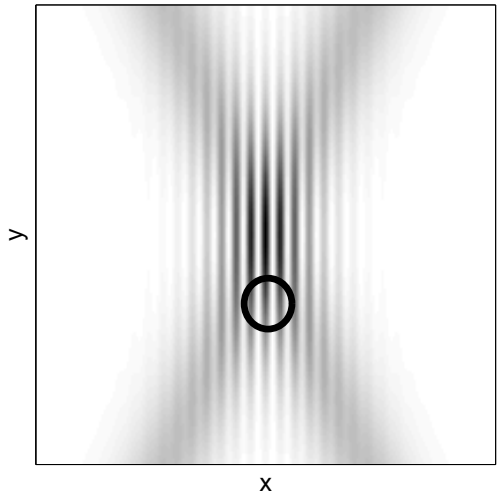}
\caption{Scheme of the proposed setup, which involves two crossing running-wave beams. Location of the 2D optical lattice created by two standing-wave laser beams (not shown) is indicated by the circle. In this setup the magnitude of artificial magnetic field (the parameter $\alpha$) is defined by the angle between the running-wave beams.}
\label{f0}
\end{figure}

The question of how one can realize the required driving in a laboratory experiment is in turn. A possible scheme, which involves two crossing running-wave beams with the wave-vector $k$, is depicted in Fig.~\ref{f0}. In this scheme we assume that (i) the laser frequencies are slightly mismatched by $\Delta\Omega$, (ii) the center of the 2D optical lattice is shifted from the crossing point  by the distance $y_0$, and (iii) there is an additional homogeneous static field of magnitude $F_0$. Then, locally, the optical potential created by two running-wave beams is given by the following expression:
\begin{equation}
\label{5}
H_{int}= F_0 (y-y_0)+ \frac{\partial V({\bf r})}{\partial y} (y-y_0) \sin^2(qx-\Delta\Omega t)  \;,
\end{equation}
where $V({\bf r})$ is proportional to the envelope function for the light intensity and  $q=k\sin\theta$. In the tight-binding approximation the potential (\ref{5}) takes the form of the  driving term in the Hamiltonian (\ref{3}) with $F=F_0+V'/2$, $F_\omega=V'/2$, $\omega=2\Delta\Omega$, and the parameter  $\alpha=(ka/2\pi) \sin\theta$. 

\section{Different gauges and phase imprinting}

\begin{figure}
\center
\includegraphics[width=14cm, clip]{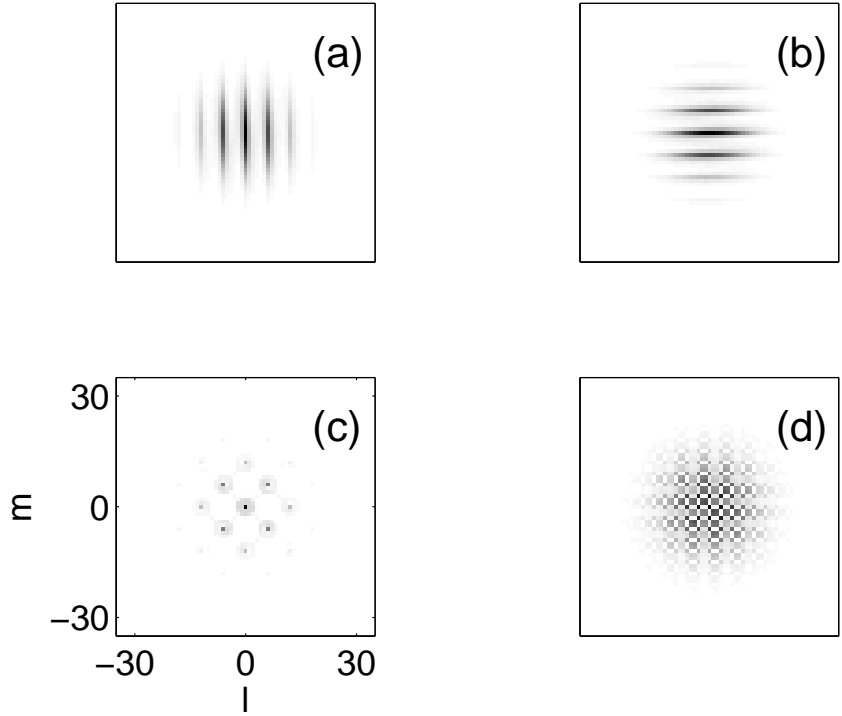}
\caption{Panels (a-c): Interference patterns at $t=10\pi$ for  $J_x=J_y=0.2$, $\alpha=1/6$, and $\gamma=2\cdot 10^{-4}$, calculated on the basis of the Hamiltonians (\ref{4a}), (\ref{4b}), and (\ref{4c}), respectively. Alternatively the figures show solution of the Schr\"odinger equation with the Hamiltonian (\ref{4a}) for the initial state (\ref{6})  with $\beta=0$, $\beta=2\pi\alpha$, and $\beta=\pi\alpha$. Panel (d): Interference  pattern for $\alpha=0$ and $\beta=\pi/6$.}
\label{f4}
\end{figure}

It is worth stressing that the proposed setup (as well as all previously suggested setups) realizes not so much a magnetic field but the tight-binding Hamiltonian of a charged particle corresponding to a given vector potential. In particular, the Hamiltonian
\begin{equation}
\label{4a}
H= -\frac{J_x}{2} \sum_l \left(|l+1,m\rangle \langle l,m |  + h.c.\right)
-\frac{J_y}{2} \sum_m \left(|l,m+1\rangle \langle m | e^{i2\pi\alpha l}  + h.c.\right) 
\end{equation}
which coincides with (\ref{4}), refers to the Landau gauge ${\bf A}=A(0,x,0)$. If we choose the gauge  ${\bf A}=A(-y,0,0)$, the Hamiltonian reads
\begin{equation}
\label{4b}
H= -\frac{J_x}{2} \sum_l \left(|l+1,m\rangle \langle l,m | e^{-i2\pi\alpha m} + h.c.\right)
-\frac{J_y}{2} \sum_m \left(|l,m+1\rangle \langle m |  + h.c.\right) \;,
\end{equation}
while for the symmetric gauge, ${\bf A}=A(-y/2,x/2,0)$, one has
\begin{equation}
\label{4c}
H= -\frac{J_x}{2} \sum_l \left(|l+1,m\rangle \langle l,m | e^{-i\pi\alpha m} + h.c.\right)
-\frac{J_y}{2} \sum_m \left(|l,m+1\rangle \langle m | e^{i\pi\alpha l}  + h.c.\right) \;.
\end{equation}
Although these Hamiltonians have the same spectrum, their eigenfuctions are different. Thus the Hamiltonians (\ref{4a}-\ref{4c}) generate different dynamics. In the other words, when evolving in time a given initial state, one gets different interference patterns.  These patterns are shown in Fig.~\ref{f4}(a-c), where we additionally assume the presence of a harmonic confinement, i.e., the Hamiltonians are augmented by the term 
\begin{equation}
\label{5b}
H_{\gamma}= \frac{\gamma}{2}\sum_{l,m}(l^2+m^2)|l,m\rangle\langle l,m| \;.
\end{equation}
As the initial state we choose the ground state of the system for $\alpha=0$,  which for $\gamma\ll 1$ is well approximated by the two-dimensional Gaussian $G(l,m)$ of the width $\sigma=(J/\gamma)^{1/4}$. In the course of time the interference patterns change with the characteristic time period defined by the cyclotron frequency, recovering quasi-periodically the initial state.

It is interesting to note that the interference patterns (b-c) can be reproduced by the system (\ref{4a}) alone, if the initial state has the imprinted phase,\footnote{In a laboratory experiment phase imprinting (\ref{6}) can be done by exposing the system to a quadrupole potential for a given time.}
\begin{equation}
\label{6}
\psi_{l,m}(t=0)=e^{-i\beta lm}G(l,m) \;,
\end{equation}
where $\beta=2\pi\alpha$ and $\beta=\pi\alpha$, respectively.  Since phase imprinting introduces an interference pattern by itself, Fig.~\ref{f4}(d) shows the case $\alpha=0$ and $\beta=\pi/6$ for the sake of comparison. At the first glance the depicted pattern resembles that for $\alpha=1/6$. However, more thorough inspection of this pattern reveals the square symmetry of the lattice, which becomes more transparent for larger times. Beside this, for $\alpha=0$, the overall width of the packet oscillates in time. This breathing of the packet width is absent if $\alpha\ne0$. It should be also mentioned that in this section we focus on the short time dynamics for $t\ll 2\pi/\Omega_\gamma$, where  $\Omega_\gamma$ is defined below in Eq.~(\ref{8}). For these times the only effect of the harmonic confinement is a modification of the boundary conditions.

We come back to the original system (\ref{3}). We have checked that the system (\ref{3}) well reproduces the interference patterns depicted in  Fig.~\ref{f4}(a-c), if condition (\ref{1c}) is satisfied. If this condition is violated,  the continuous dynamics of the driven system (\ref{3}) may not coincide with dynamics of the target system (\ref{4}). However, even in this case the discrete dynamics of (\ref{3}) and (\ref{4}) do coincide.\footnote{By discrete dynamics we mean the dynamics in terms of the system evolution operator over the driving period.}

\section{Cyclotron dynamics}

Observation of the interference patterns discussed in the previous section requires $\sigma\gg 1/\alpha$. In this section we analyze the opposite situation, where the magnetic length $1/\alpha$ is larger than the wave packet width $\sigma$. In this case the wave function does not develop an interference pattern. Nevertheless, one can detect the artificial magnetic field by watching the evolution of the wave-packet center of gravity, which is expected to follow the classical trajectory of the system. Needless to say that the classical trajectories (as well as the wave-packet dynamics associated with these trajectories) do not depend on the gauge.\footnote{For plane lattices (no harmonic confinement) the wave-packet dynamics of the system has been studied in the recent paper \cite{preprint}. It is also shown there that the semiclassical regime of the wave-packet dynamics corresponds to the limit $\alpha\rightarrow0$, where the cyclotron radius essentially exceeds the lattice period.} 

\begin{figure}
\center
\includegraphics[width=12cm, clip]{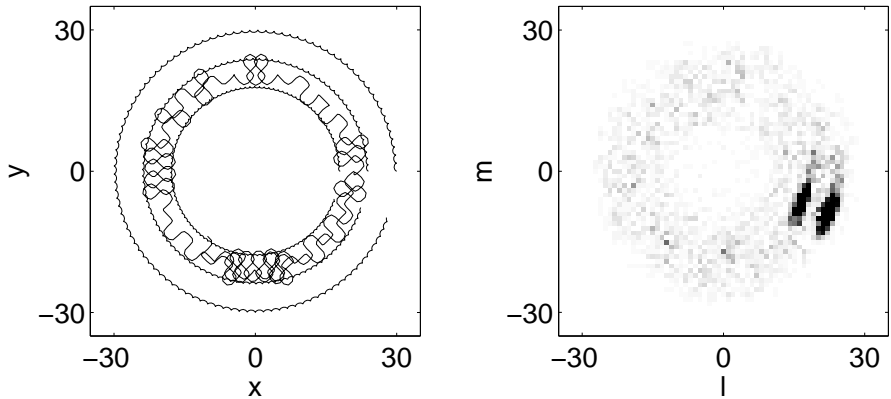}
\caption{Left: 4 different classical trajectories for $\alpha=1/6$, $J_x=J_y=1$ and $\gamma=0.01$. Computational time is $t=200\pi$, which approximately coincides with $T_\gamma=209.44\pi$. Right: Occupation probabilities at $t= 200\pi$. Initial wave packet corresponds to the ground state of the system for $\alpha=0$, shifted by 21 lattice periods to the right.}
\label{f6}
\end{figure}

The classical counterpart of the quantum tight-binding Hamiltonian (\ref{4a}) reads \cite{preprint}
\begin{equation}
\label{7}
H_{cl}=-J_x\cos p_x - J_y\cos(p_y+2\pi\alpha x) +\frac{\gamma}{2}(x^2 + y^2) \;,
\end{equation}
where the last term takes into account the parabolic confinement and we set the lattice period $a$ to unity. As initial conditions we consider $p_x=p_y=0$, and $(x,y)=(x_0,y_0)$. In a laboratory experiment one realizes these initial conditions by suddenly shifting the center of the lattice by the distance ${\bf r}=(x_0,y_0)$ \cite{Cata03}. For $\alpha=0$ this shift induces dipole oscillations, where the wave-packet center of gravity oscillates around the lattice origin with the frequency $\Omega_{d}=\sqrt{J\gamma}/\hbar$ \cite{Cata03,dipole}. Unlike these dipole oscillations, for $\alpha\ne0$ the classical trajectory encircles the lattice origin (see Fig.~\ref{f6}). Note that for $\alpha\ne0$ the Hamiltonian (\ref{7}) is not separable and, hence, trajectories may be regular or chaotic. Regular trajectories correspond to the stability islands around the points, where arguments of the cosine functions in (\ref{7}) are multiple of $2\pi$ and where one may use the effective mass approximation. These trajectories encircle the center of the parabolic lattice in the same direction (defined by the sign of $\alpha$) with the frequency
\begin{equation}
\label{8}
\Omega_\gamma=\gamma/2\pi\hbar\alpha \;.
\end{equation}
Chaotic trajectories have no well defined encircling direction and change it in a random way. This observation helps us to understand the result of numerical simulation of the wave-packet dynamic shown in the right panel in Fig.~\ref{f6}. Here the main part of the wave packet rotates counterclockwise with the frequency (\ref{8}), while a fraction of the packet goes in the opposite direction. Concluding this section we note that the frequency (\ref{8}) can be also obtained quantum-mechanically, -- it corresponds to the splitting  $\Delta E=\hbar\Omega_\gamma$ of the low-energy Landau levels of the systems (\ref{4a}-\ref{4c}) caused by the parabolic term (\ref{5b}).

\section{Conclusion}

We have proposed a simple setup for introducing an artificial magnetic field for neutral atoms in 2D optical lattices. This setup involves a low-frequency driving of the optical lattice, which can be easily realized by the same means which employed to create the lattice. Note that such low-frequency driving does not affect the electronic structure of the atom. This constitutes the main difference of our setup from the previously suggested setups, which involve Raman transitions between atomic spin states to mimic the magnetic field \cite{Jaks03,Muel04,Oster05,Lin09,Sore05,Juze04,Ruse05,Guen09}.

In the second part of the paper we have discussed two dynamic-based methods of detecting the (artificial) magnetic field. Importantly, we take into account harmonic confinement which is inevitably present in a laboratory experiment. 

The first method is the observation of interference patterns developed by the atomic wave function in the course of time. This method assumes a highly coherent initial state (a Bose-Einstein condensate) with the coherence length exceeding the magnetic period. A variety of interference patterns signaling the presence  of the effective magnetic field is identified.

The second method does not require a coherent initial state. It relies on the underlying classical dynamics of an atom in a parabolic lattice. For vanishing magnetic field these dynamics consist of dipole oscillations with the frequency defined by the strength of  the parabolic confinement \cite{dipole}. Unlike these dipole oscillations, for a finite field the atom encircles the center of the parabolic lattice with a frequency defined by both the strength of  the parabolic confinement and the magnetic field magnitude. This `encircling frequency' is shown to be well seen in the quantum wave-packet dynamics. Thus the artificial magnetic field can be measured by measuring the encircling frequency.  

\acknowledgements
This work was supported by Russian Foundation for Basic Research, grant RFBR-10-02-00171-a.

\end{document}